\documentclass[usenatbib]{mn2e}
\usepackage{multirow}
\usepackage{rotating}
\usepackage{colortbl}
\usepackage{color}
\usepackage[fleqn]{amsmath}
\usepackage{amssymb}
\usepackage{amsfonts}
\usepackage{verbatim}
\usepackage{scalefnt}
\usepackage[percent]{overpic}

\newcommand{\kms}{\,{\rm km\,s^{-1}}}
\newcommand{\msun}{\,{\rm M_\odot}}

\newcommand{\edd}{{\rm Edd}}

\newcommand{\beq}{\begin{equation}}
\newcommand{\eeq}{\end{equation}}
\newcommand{\ba}{\begin{eqnarray}}
\newcommand{\ea}{\end{eqnarray}}
\def\spose#1{\hbox to 0pt{#1\hss}}
\newcommand{\lta}{\mathrel{\spose{\lower 3pt\hbox{$\mathchar"218$}}
      \raise 2.0pt\hbox{$\mathchar"13C$}}}
\newcommand{\gta}{\mathrel{\spose{\lower 3pt\hbox{$\mathchar"218$}}
      \raise 2.0pt\hbox{$\mathchar"13E$}}}

\newcommand{\comments}[1]{} %usage: \comments{}
\title[Appearance of supercritical BHs]{Hyperaccreting black holes in galactic nuclei}

\author[Begelman  and Volonteri]{Mitchell C. Begelman,$^{1,2}$ Marta Volonteri$^{3}$\\
$^{1}$JILA, 440 UCB, University of Colorado at Boulder, Boulder, CO 80309-0440, USA\\
$^{2}$ Department of Astrophysical and Planetary Sciences, 391 UCB, University of Colorado, Boulder, CO 80309-0391, USA\\
$^{3}$ Institut d'Astrophysique de Paris, Sorbonne Universit\'es, UPMC Univ. Paris 06 et CNRS, UMR 7095, F-75014, Paris, France}

\begin{document}

\maketitle

\begin{abstract}
The rate at which matter flows into a galactic nucleus during early phases of galaxy evolution can sometimes exceed the Eddington limit of the growing central black hole by several orders of magnitude.  We discuss the necessary conditions for the black hole to actually accrete this matter at such a high rate, and consider the observational appearance and detectability of a hyperaccreting black hole.  In order to be accreted at a hyper-Eddington rate, the infalling gas must have a sufficiently low angular momentum.  Although most of the gas is accreted, a significant fraction accumulates in an optically thick envelope with luminosity $\sim L_\edd$, probably pierced by jets of much higher power.  If $\dot M > 10^3 M_\edd$, the envelope spectrum resembles a blackbody with a temperature of a few thousand K, but for lower (but still hyper-Eddington) accretion rates the spectrum becomes a very dilute and hard Wien spectrum.  We consider the likelihood of various regimes of hyperaccretion, and discuss its possible observational signatures.
\end{abstract}


\section{Introduction}\label{sec:Introduction}

Episodes of hyperaccretion --- accretion at rates far exceeding the Eddington limit --- are often invoked to explain the early rapid growth of massive black holes \citep[MBHs:][and references therein]{volonteri05,volonteri15}.  Theoretical estimates of mass supply rates available in protogalaxies are certainly compatible with hyperaccretion.  For example, the characteristic infall rate of self-gravitating gas in a halo with velocity dispersion $100 \sigma_{100}\kms$,
\beq
\dot M_{\rm ff} = {\sigma^3\over G} = 240 \sigma_{100}^3 \msun \ {\rm yr}^{-1},
\label{msg}
\eeq 
exceeds the Eddington rate for a $10^6 m_6 \msun$ black hole by a factor $\sim 10^5 \sigma_{100}^3 m_6^{-1}$. Here we have defined the Eddington accretion rate assuming electron scattering opacity and without an overall radiative efficiency, i.e., $\dot M_\edd = L_\edd / c^2$.   

Observational surveys suggest that the distribution of mass supply rates in active galactic nuclei (AGN), normalized to the Eddington value, is mass-independent, and is a decreasing function of the Eddington ratio but not a steep one, e.g., a power-law with index $\sim -0.65$ \citep{aird12}, which \cite{aird13} suggest to steepen to $\sim -2$ or cut off with an exponential \citep{stanley15} for Eddington ratios larger than unity. If we instead extrapolated these power-laws to high accretion rates we would infer a non-negligible fraction of supercritical AGN, $\sim 10^{-3}$ at $z=1$ and $\sim 10^{-2}$ at $z=2$. These fractions, however, are derived from extrapolation of results for low-redshift ``normal" AGN.   For gas-rich protogalaxies at high redshifts, particularly following mergers, or for black holes that have not yet grown to their final masses, the occasional availability of gas at a hypercritical rate is more likely, as the same fraction of $\dot M_{\rm ff}$ represents a higher fraction of $\dot M_\edd$ for a lower-mass black hole. Indeed, large-scale simulations of galaxy assembly suggest that supercritical infall rates are fairly common \citep[e.g.,][]{dubois14}.

The availability of gas at a hypercritical rate is a separate question from the acceptance of such gas by the black hole; the latter is the subject of much uncertainty.  X-ray binaries with hyper-Eddington mass transfer rates, such as SS 433 \citep{begelman06}, microquasars in outburst \citep{neilsen16} and ultraluminous X-ray sources \citep{poutanen07,middleton15,pinto16}, appear to eject much of the supplied gas before it reaches the black hole. This regulates the accretion luminosity to a moderately supercritical value, as posited in the inflow-outflow models of \cite{shakura73} and \cite{blandford99}.  These systems have in common that the mass is supplied through a thin disk, with nearly Keplerian angular momentum.  Only after the gas passes the trapping radius, $r_{\rm tr} = (\dot M / \dot M_\edd) r_{\rm g}$ \citep{begelman79}, where $r_{\rm g}= GM/c^2$ is the black hole's gravitational radius, does radiation energy density build up in the flow, thickening the disk and apparently driving the outflow.  The mechanism driving the outflow in specific cases is not well understood, but it could be that the combination of strong angular momentum transport, in conjunction with relatively weak photon trapping, allows enough gas to escape at each radius that the radiation remains marginally trapped all the way in to the center.  Although the outer geometry of accretion flows in AGN is not well known, it is possible that high angular momentum disk accretion similarly occurs there, so that few if any AGN produce hypercritical luminosities.

The situation appears to be different in a small subset of candidate tidal disruption events (TDEs), where debris from the disrupted star falls back toward the black hole at a hypercritical rate and appears to be accreted without difficulty, liberating a hypercritical luminosity and powering jets \citep{zauderer11,cenko12,brown15}.  A key difference between this case and the cases with strong mass loss and regulated accretion is that the specific angular momentum of the infalling gas is far below the Keplerian angular momentum at the trapping radius.  The radiation is therefore strongly coupled to the gas where the angular momentum is deposited, and as a result it may not be able to drive strong mass loss.  

To describe this case, \cite{coughlin14} proposed that instead of being blown away, the infalling gas would inflate into a weakly bound envelope which they called a ZEBRA (ZEro-BeRnoulli Accretion flow).  Essentially all the matter in such an envelope could be accreted if the gas close to the black hole could be pushed into low binding-energy relativistic orbits before falling in, as in the ``Polish Doughnut" model \citep{jaroszynski80}, or finds a way of venting most of its accretion energy into the centrifugally evacuated funnel around the rotation axis, as seems to happen in the candidate TDEs.  In either case, the luminosity leaking out in directions other than the accretion funnel would be limited to roughly $L_\edd$.

In this paper we adopt the view that hyperaccretion is bimodal: either most of the gas is blown away and the residual accretion rate is close to Eddington, or the matter is accreted at nearly the rate supplied.  The parameter that discriminates between the two cases is the ratio of the specific angular momentum in the supplied gas to its Keplerian value at the trapping radius, which is a function of the mass supply rate, $\dot M$.  

In section 2 we summarize the parameters that determine the outcome of hyperaccretion, and in section 3 we study the structure and evolution of the ZEBRA envelope expected to develop in the low angular momentum case, expressing our results in terms of the fraction of black hole mass acquired during hyperaccretion episodes.  We consider the appearance of highly supercritical black holes in section 4, and discuss the observational consequences in section 5.   We summarize our results and conclude in section 6.

\section{Parameter space for hyperaccretion}\label{sec:sec2}

Assume that matter is supplied to the black hole (mass $M$) at a rate $\dot M$, measured at an outer accretion radius which we define as $r_{\rm B} = GM/\sigma^2 $, by analogy with the Bondi radius.  This is also the black hole's radius of influence in the galactic nucleus potential.  For various purposes we will choose to normalize $\dot M$ to $\dot M_{\rm ff}$, $\dot {\tilde m} = \dot M / \dot M_{\rm ff}$, or to $\dot M_\edd$, $\dot m = \dot M/ \dot M_\edd$. Likewise, we define scaled radii normalized to $r_{\rm B}$, $\tilde x = r/r_{\rm B}$, and to $r_{\rm g}$, $x = r/r_{\rm g}$. Scaled to $r_{\rm B}$, the trapping radius can be written
\beq
\label{rtrB}
\tilde x_{\rm tr} = \dot m \left({\sigma \over c}\right)^2 \approx 10^{-7} \dot m \sigma_{100}^2 \approx 10^{-2} \tilde\kappa \dot {\tilde m} \sigma_{100}^5 m_6^{-1}.
\eeq
Now define a dimensionless angular momentum parameter $\lambda_{\rm B} = \ell_{\rm B} /(GMr_{\rm B})^{1/2}$, where $\ell_{\rm B}$ is the specific angular momentum of gas at the accretion radius.  We identify several regimes of the accretion process depending on the value of $\lambda_{\rm B}$:

\begin{itemize}
  \item $\lambda_{\rm B} \geq 1$. The accretion flow is completely regulated by angular momentum transport and ultimately transitions to an inflow-outflow state, as in a hyperaccreting X-ray binary. 

\item $\tilde x_{\rm tr}^{1/2} < \lambda_{\rm B} < 1 $. The gas falls through the accretion radius without feeling centrifugal effects, but encounters a centrifugal barrier before reaching the trapping radius.  As a result, the stalled gas cools and forms a disk at the Keplerian radius.  According to our assumption about the bimodal character of hyperaccretion, this implies that the flow undergoes strong mass loss inside the trapping radius, leaving the black hole with only mildly supercritical accretion.

\item $3\times 10^{-4} \sigma_{100} < \lambda_{\rm B} < 0.1 \dot{\tilde m}^{1/2} \sigma_{100}^{5/2} m_6^{-1/2} \equiv \lambda_{\rm B, crit}$.  This is the most interesting case from the perspective of this paper.  The lower limit implies that the gas has too much angular momentum to fall directly into the black hole, and thus must dissipate some binding energy first.  This condition is expected to be satisfied in the vast majority of cases.  The upper limit implies that the gas reaches its centrifugal barrier after it has fallen through the trapping radius.  This means that angular momentum is deposited under highly trapped conditions, which we have speculated leads to a weakly bound envelope (a ZEBRA flow) with little mass loss, feeding the black hole at a hypercritical rate.

\end{itemize}

\section{Structure and evolution of a hyperaccreting envelope}\label{sec:sec3}

The structure and evolution of ZEBRA flows is discussed by \cite{coughlin14}, in the context of hypercritical fallback rates following TDEs. As in that case, the black hole is expected to swallow most of the matter falling back in real time while absorbing little of the angular momentum, which resides mainly in the outer regions of the flow. The main difference between the case considered here and the TDE case is that $\dot M$ is roughly constant with time over the course of the hyperaccretion event, whereas in the TDE case the fallback rate declines $\propto t^{-5/3}$.  As a result, in the TDE case most of the angular momentum that needs to be stored is accumulated during the early stages of fallback, after which the stored angular momentum, and envelope mass, remain roughly constant. For the case considered here, the envelope must contain an increasing amount of gas in order to store the leftover angular momentum, until it can be transferred to the environment.     

Initially, the part of the accretion flow below the trapping radius contains enough mass to absorb the accumulated angular momentum without expanding appreciably. The initial mass in the trapped region of the accretion flow is 
\beq
\label{mtr}
M_{\rm tr} \approx 10^3 \dot{\tilde m}^{5/2} \sigma_{100}^{15/2} m_6^{-1/2} \ \msun, 
\eeq     
which can handle the angular momentum deposited by the accretion of mass $M_{\rm acc} = \dot M t = M_{\rm tr} (\lambda_{\rm B, crit}/\lambda_{\rm B} )$.  This capacity is exceeded very quickly, after an elapsed time
\beq
\label{mtr}
t_{\rm tr} \approx 4 \dot{\tilde m}^{3/2} \sigma_{100}^{9/2} m_6^{-1/2} \left({\lambda_{\rm B, crit}\over \lambda_{\rm B} } \right) \ {\rm yr}, 
\eeq     
and the hyperaccreting black hole must then begin to accumulate a more massive gaseous envelope to carry the deposited angular momentum.  

The trapping condition, which we have expressed in terms of a steady accretion rate, can also be interpreted in terms of an envelope mass $M_{\rm env} = m_{\rm env} \msun$ which is forced to convect radiation with a characteristic speed approaching $\sim v_{\rm K}$, the local Keplerian speed \citep{coughlin14}, but participates in the accretion flow at a slower rate.  Equivalently, the trapping condition can be obtained by setting the characteristic optical depth across the trapping radius equal to $c/ v_{\rm K}$.  The resulting envelope radius is given by 
\beq
\label{renv}
r_{\rm env} \approx 8.8 \times 10^{14} \tilde\kappa^{2/5} m_{\rm env}^{2/5} m_6^{1/5} \ {\rm cm}
\eeq 
\citep{begelman12b}, where $\tilde \kappa \equiv \kappa / \kappa_{\rm es}$ is the opacity normalized to electron scattering opacity.  (We include $\tilde\kappa$ here and not in the expression for $r_{\rm tr}$ above because the envelope may become much more extended and cooler than $r_{\rm tr}$, and thus be subject to different opacity.) Simultaneously, the envelope must be large enough to contain the total angular momentum ${\cal L}_{\rm env}$, which requires (to within a constant of order unity) ${\cal L}_{\rm env} \sim M_{\rm env} (GMr_{\rm env})^{1/2}$ or, equivalently,
\beq
\label{LMenv}
M_{\rm env} \approx  \left({4\pi c\over \kappa} \right)^{1/6} {\cal L}_{\rm env}^{5/6} (GM)^{-1/2} 
\eeq 
(Coughlin \& Begelman 2014).

Neglecting the angular momentum lost through accretion and winds or jets, we can relate ${\cal L}_{\rm env}$ to the (assumed steady) accretion rate,
\beq
\label{Lenv}
\begin{split}
{\cal L}_{\rm env} &= \lambda_{\rm B} \dot M t (GM r_{\rm B})^{1/2} \\ 
&\approx  6 \times 10^{59} \left({\lambda_{\rm B}\over \lambda_{\rm B, crit} } \right) \dot{\tilde m}^{3/2} \sigma_{100}^{9/2}m_6^{1/2} t_{\rm yr} \ {\rm g\ cm}^2 \ s^{-1},
\end{split}
\eeq 
where $t_{\rm yr}$ is the elapsed time of the hyperaccretion episode in years. The mass of the envelope increases according to
\beq
\label{Menv}
M_{\rm env} \approx 240 {\tilde\kappa}^{-1/6} \left({\lambda_{\rm B}\over \lambda_{\rm B, crit} } \right)^{5/6} \dot{\tilde m}^{5/4} \sigma_{100}^{15/4} m_7^{-1/12} t_{\rm yr}^{5/6} \ \msun,
\eeq 
for $t > t_{\rm tr}$.

The steady increase of envelope mass should continue until either the episode of hyperaccretion ends due to a decrease in the mass supply for accretion, or some mechanism sets in that removes the excess angular momentum and/or accumulating mass.  One  plausible mechanism for promoting angular momentum loss would be the self-gravity of the envelope, which becomes important when the envelope mass approaches that of the black hole.  However, one can see from equation (\ref{Menv}) that this limit is never reached, because the envelope mass grows more slowly than the black hole mass.  For example, in an extreme case where the black hole grows by more than an order of magnitude during a single episode of steady hyperaccretion, we see that $M \propto t$ while $M_{\rm env} \propto t^{3/4}$ (because $dM_{\rm env} /dt \propto M^{-1/12} t^{-1/6}$).

If the maximum mass reached by the envelope is limited by the duration of the hyperaccretion episode, it is instructive to characterize this in terms of the fraction $f$ of the current black hole mass acquired during $N$ episodes of hyperaccretion, all of which we assume to be similar.  We assume that the envelope is drained between episodes.  The duration of the current hyperaccretion episode is then given by $t_{\rm hyp} = (f/N)(M/ \dot M)$, and the mass reached by the envelope during one episode is 
\beq
\label{Mhyp}
M_{\rm hyp} \approx 3 \times 10^5 {\tilde\kappa}^{-1/6} \left({\lambda_{\rm B}\over \lambda_{\rm B, crit} } {f\over N}\right)^{5/6}  \dot{\tilde m}^{5/12} \sigma_{100}^{5/4} m_6^{3/4}  \ \msun.
\eeq 
The associated radius, from equation (\ref{renv}), is
\beq
\label{rhyp}
r_{\rm hyp} \approx 1.4 \times 10^{17} {\tilde\kappa}^{4/15} \left({\lambda_{\rm B}\over \lambda_{\rm B, crit} } {f\over N}\right)^{1/3}  \dot{\tilde m}^{1/6} \sigma_{100}^{1/2} m_6^{1/2}  \ {\rm cm}.
\eeq 

We now consider the density distribution inside the envelope, which is essential for calculating its spectral properties.  The mass and outer radius of the envelope (equivalently, the trapping condition) fix the characteristic density of the outer envelope.  The inner density, at a few gravitational radii, is fixed by our assumption that matter is being fed into the black hole roughly at the rate it is being supplied, $\dot M$.  Although we are ignorant of the detailed distribution of angular momentum within the envelope, we note that it must approach the local Keplerian value at both inner and outer radii.  Therefore, a simple assumption to connect the two zones is that the specific angular momentum distribution is quasi-Keplerian, i.e., scaling according to $\ell^2 \sim  a GMr$, where $a < 1$ is a constant and we are using the notation of \cite{coughlin14}, who showed that the value of $a$ and the slope of the density profile, $q \equiv - d\ln \rho/d\ln r$, are interrelated. 

To estimate the required slope, we assume that matter contained within an inner radius $r_0 \sim 10 r_{\rm g}$ is falling into the black hole at a speed $v_0 \sim 0.03 c$.  This implies that the mass contained within $r_0$ is roughly $M_0 = \dot M r_0/v_0 = 0.01 \dot{\tilde m}\sigma_{100}^3 m_6 \ \msun$.  The corresponding density is obtained from solving $\dot M = 4\pi \rho_0 v_0 r_0^2$.  We then have
\beq
\label{qest}
3 - q \approx {\ln (M_{\rm hyp}/M_0)\over \ln (r_{\rm hyp}/r_0)} \approx 1.5 {1 + 0.06 \ln A\over 1+ 0.09 \ln B},
\eeq   
where 
$A = {\tilde\kappa}^{-1/6} [(\lambda_{\rm B}/\lambda_{\rm B, crit}) (f/ N)]^{5/6}  \dot{\tilde m}^{-7/12} \sigma_{100}^{-7/4} m_6^{-1/4}$, $B = {\tilde\kappa}^{-4/15} [(\lambda_{\rm B}/\lambda_{\rm B, crit}) (f/ N)]^{1/3}  \dot{\tilde m}^{1/6} \sigma_{100}^{1/2} m_6^{-1/2}$. For a wide range of plausible parameters we can neglect the log terms and take $3- q\approx 1.5$, which gives a density slope very close to that of free-fall.   To simplify subsequent expressions, given the crudity of our approximations so far, we will take $q = 3/2$.  The density is then given by 
\beq
\label{rhohyp}
\rho \approx 1.9 \times 10^{-5} \dot{\tilde m} \sigma_{100}^3 m_6^{-2} x^{-3/2}  \ {\rm g \ cm}^{-3},
\eeq 
where we recall that $x = r/r_{\rm g}$.

\section{Radiative properties of hyperaccreting black holes}\label{sec:sec4}

If the envelope formed a photosphere close to $r_{\rm hyp}$, its effective temperature, assuming $L= L_\edd$, would be
\beq
\label{Teff}
T_{\rm eff} \approx 1.8 \times 10^{3} {\tilde\kappa}^{-23/60} \left({\lambda_{\rm B}\over \lambda_{\rm B, crit} } {f\over N}\right)^{-1/6}  \dot{\tilde m}^{-1/12} \sigma_{100}^{-1/4}  \ {\rm K},
\eeq 
which is insensitive to all the parameters.  Scattering in the outer layers could increase the color temperature over $T_{\rm eff}$ by a factor 2 or 3; the high temperature sensitivity of H$^-$ opacity would prevent the photospheric temperature from dropping below a few thousand K.    

However, the relatively low density and higher temperature close to the black hole suggest that under certain circumstances, thermalization may fail outside a relatively small radius compared to $r_{\rm hyp}$, in which case the spectrum will resemble a dilute blackbody with a color temperature much higher than $T_{\rm eff}$.  This was pointed out by \cite{beloborodov98} in the context of slim disk models for hyperaccretion, where most of the liberated energy is assumed to be advected into the black hole. 

To assess the level of thermalization, we assume that the interior of the envelope is electron scattering-dominated, with an absorption opacity given by the standard Kramers formula for bound-free absorption, $\kappa_{\rm bf} \approx 1.6 \times 10^{24} \rho T^{-7/2}$. The (radiation) pressure in the envelope is given by 
\beq
\label{phyp}
p \approx {2\over 11} \rho {GM\over r} \approx 3.1 \times 10^{15} \dot{\tilde m} \sigma_{100}^3 m_6^{-2} x^{-5/2}  \ {\rm erg \ cm}^{-3},
\eeq    
corresponding to an LTE temperature
\beq
\label{Thyp}
T_{\rm LTE} \approx 3.5 \times 10^{7} \dot{\tilde m}^{1/4} \sigma_{100}^{3/4} m_6^{-1/2} x^{-5/8}  \ {\rm K}.
\eeq    
The effective optical depth for thermalization as a function of $r$ is then given by $\tau_{\rm LTE} = \rho (\kappa_{\rm bf} \kappa_{\rm es})^{1/2} r \propto r^{-5/32}$, and the radiation is thermalized at radii smaller than $r_{\rm LTE}$, where $\tau_{\rm LTE} (r_{\rm LTE}) = 1$. It turns out that the thermalization radius is extremely sensitive to $\dot M$, varying more steeply than $\propto \dot M^6$. The interior first starts to lose thermalization when the accretion rate drops below
\beq
\label{mdotth}
\dot{\tilde m}_{\rm th} \approx 0.016 {\tilde\kappa}^{0.04} \left({\lambda_{\rm B}\over \lambda_{\rm B, crit} } {f\over N}\right)^{0.05}  \sigma_{100}^{-3} m_6 ,
\eeq
corresponding to 
\beq
\label{Mdotth}
\dot m \equiv {\dot M \over \dot M_\edd} \lesssim 1.4 \times 10^3 .
\eeq
The thermalization radius decreases rapidly with decreasing $\dot M$ until the entire accretion flow becomes unthermalized (i.e., down to radii $r_0\sim 10 r_{\rm g}$), for $\dot M$ about four times smaller than $\dot{\tilde m}_{\rm th}$, i.e., for $\dot M \sim 350 \dot M_\edd$.  Note that these limits depend almost entirely on conditions in the accretion flow at the inner radii, $\sim r_0$, and therefore are insensitive to the uncertain outer structure of the envelope.  However, as \cite{beloborodov98} shows, the estimated temperatures are likely to to be quite sensitive to the inner conditions mainly through the density and emitting volume, and thus one can obtain considerably higher temperatures for larger inflow speeds (i.e., larger viscosity parameter $\alpha$) and more rapidly spinning black holes.     

When the inner flow is just barely thermalized, the LTE temperature at $r_0$ is $\sim 2 \times 10^6 m_6^{-1/4}$ K, and for lower $\dot M$ the radiation is presumably supplied by Comptonized bremsstrahlung, with a Wien spectrum and a nominal temperature of  
\beq
\label{Tth}
T \approx 2.1 \times 10^6  m_6^{-1/4} \left({\dot M\over 350 \dot M_\edd} \right)^{-4} \ $\rm K$,
\eeq
which in practice will be depressed by a logarithmic factor due to Comptonization effects \citep{rybicki79}.  We note that this temperature rapidly approaches and can exceed the virial temperature in the outer parts of the accretion flow, raising the possibility that Compton pre-heating could quench the flow \citep{inayoshi16}.  However, as we argue below, the most likely cases of hyperaccretion occur at such high accretion rates that the radiation is thermalized and therefore emerges in the optical and infrared.

\begin{figure}
\begin{center}
\includegraphics[width=\columnwidth]{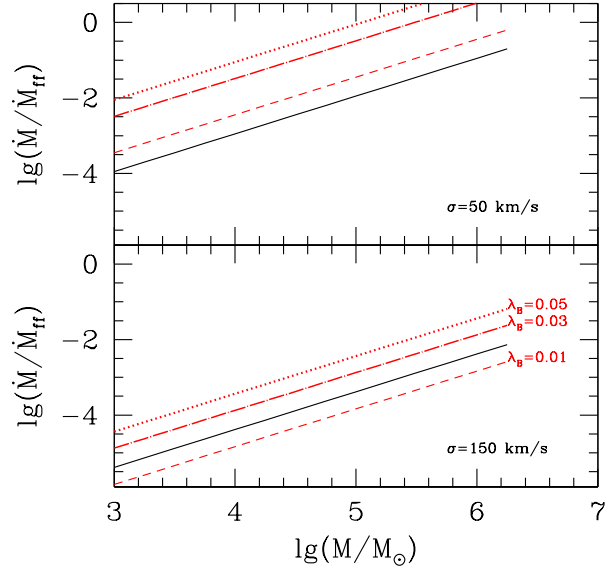}
\caption{Examples of conditions leading to a RedZEBRA or an XZEBRA. The accretion flow around a black hole is a ZEBRA above the red dashed, dot-dashed and dotted lines, for various values of $\lambda_{\rm B}$, as a function of black hole mass and $\dot{\tilde m}$ in a galaxy with a given $\sigma$ (50 or 150 $\kms$ in this example). Above the black solid line it would appear as a RedZEBRA, while below it would appear as an XZEBRA. }
\label{fig:zebras}
\end{center}
\end{figure}

\section{Observational consequences}\label{sec:sec5}
The most favorable conditions for MBHs to become ZEBRAs are associated with the largest values of $\lambda_{\rm B, crit}$, i.e., a relatively small black hole in a relatively massive galaxy with a large $\sigma$ (cf.~section 2). If the growth of the black hole is ``left behind," so that the galaxy grows faster than the black hole, even a small fraction of infalling gas with sufficiently low angular momentum can give rise to a ZEBRA episode. For instance, if $m_6=0.1$ and $\sigma_{100}=1$, a very modest 1\% of gas at free-fall rate needs to have low angular momentum, $\lambda_{\rm B}\sim 0.03$, to trigger an accretion episode with $\dot m\sim 10^4$. 

Equation (\ref{Mdotth}) implies that there are two different regimes of hyperaccreting black holes in the ZEBRA mode.   The quasi-isotropic emission at $\sim L_\edd$ of ZEBRAs will resemble that of a red giant if the following joint condition is met:
\beq
\label{redzebra}
\dot{\tilde m}>0.014 \frac{m_6}{\sigma_{100}^{3}} \ \ {\rm and} \ \ \dot{\tilde m}  >  \left(\frac{\lambda_{\rm B}}{0.1}\right)^2 {m_6}{\sigma_{100}^{-5}} .
\eeq
We call this a RedZEBRA.  For a low-mass black hole ($\sim 10^5 \msun$), gas needs to flow in at a very small fraction of the free-fall rate ($\dot{\tilde m}\sim 10^{-3}$) to fulfill this criterion. 

Alternatively, if 
\beq
\left(\frac{\lambda_{\rm B}}{0.1}\right)^2  {m_6}{\sigma_{100}^{-5}}< \dot{\tilde m}  < 0.014 \, {m_6}{\sigma_{100}^{-3}},
\eeq
the ZEBRA will be a hard X-ray source, an XZEBRA.  For instance, in a galaxy with $\sigma=150 \,\kms$ and a black hole with $m_6=1$, for gas with $\lambda_{\rm B}=0.01$, we would have an XZEBRA if the inflow rate is $1.5\times 10^{-3}<\dot{\tilde m}<4\times 10^{-3}$ and a RedZEBRA if $\dot{\tilde m}>4\times 10^{-3}$. In a galaxy with $\sigma=50 \,\kms$, no XZEBRAS can occur if $\lambda_{\rm B}> 0.005-0.006$.  We summarize these constraints in Fig.~1.

 RedZEBRAs would appear as very luminous red sources, with a luminosity of $L\sim 10^{44} {m_6} \, {\rm erg \, s^{-1}}$ peaking at $1.6 \mu m$ rest-frame for a temperature of $T_{\rm eff} \approx 1.8 \times 10^{3}$ K. In principle, the redshifted blackbody peak would be accessible to the {\it James Webb Space Telescope} ({\it JWST}) out to $z\sim2$ with NIRCAM and out to $z\sim 16.5$ with MIRI. XZEBRAs are also within the reach of future X-ray telescopes, such as {\it Athena}, or very deep fields with current instruments. Several factors, however, may limit detectability. One is the likely short duration of hyperaccretion events, which could not be sustained for much more than $\sim 10^5$ yr at $\dot m \sim 10^3-10^4$, even if the black hole acquires most of its mass through hyperaccretion.  A second factor is that hyperaccretion might be quenched by feedback after a fairly short time \citep{volonteri15}.  A third factor, in the case of XZEBRAs, is obscuration due to the material feeding the black hole, which \cite{volonteri15} estimate within the Bondi radius as
\beq
N_H\sim \dot{\tilde m}  \frac{\sigma^4}{m_p G M}\sim  10^{26} \dot{\tilde m} m_6^{-1}\sigma_{100}^4  \, {\rm cm}^{-2}.
\label{NH}
\eeq
The column density is dominated by material close in, and if the outer material remains cool, downscattering of the hard X-rays may still contribute to the emission.  

In addition to the quasi-isotropic emission at $\sim L_\edd$, either kind of ZEBRA may have powerful jets emerging along the rotational axis, carrying a power $\sim \epsilon \dot m L_\edd$, where $\epsilon$ is the accretion efficiency.  The emission looking down the axis of one of these jets may be the most efficient way to detect hyperaccreting MBHs at high redshifts. By analogy with the super-Eddington TDE Swift J1644+57, geometric beaming rather than relativistic Doppler shift may be the most important factor in enhancing the apparent luminosity over the actual luminosity, and the detected radiation may represent mildly relativistic outflow along walls of the accretion funnel \citep{kara16}. The spectrum would thus resemble that in the inner regions of the accretion flow, i.e., soft- to medium-energy X-rays for the RedZEBRA cases, ranging to extremely hard X-rays for XZEBRAs.  If the beaming factors are as large as they apparently are in TDEs ($b\sim 10^2$ for Swift J1644+57) the apparent X-ray luminosity of a hyperaccreting $10^5$ or $10^6 \msun$ black hole could exceed those of the most luminous quasars, but at the cost of only a small fraction of the sources being visible in this way.  An extended radio jet is not likely at $z>4$, because the relativistic electrons cool preferentially by scattering cosmic microwave background photons, rather than by synchrotron emission. The radio lobes would then be quenched at high redshifts, but compact hotspots could still be visible at low frequencies \citep{2015MNRAS.452.3457G}. Given the additional fact that accretion rates of $\dot m \sim 10^3 - 10^4$ can only be sustained for $\lta 10^5$ yr before the black hole grows out of its mass range, it is likely that these sources are quite rare.

\section{Summary and conclusions}

We have considered the conditions under which massive black holes in galactic nuclei might accept infalling matter at an extremely super-Eddington rate, and how such hyperaccreting MBHs might be detectable.  We adopt a bimodal criterion for hyperaccretion, in which the black hole is able to swallow material at a hypercritical rate only if this matter falls within the radiation trapping radius without first forming a disk.  This argument is based on observations of SS 433 and other X-ray binaries undergoing mass transfer at a hypercritical rate, where there is evidence that powerful winds expel most of the supplied mass before it reaches the black hole, in contrast to hyperaccreting TDEs \citep{coughlin14}, where the infalling gas has a very low angular momentum compared to the Keplerian value at the trapping radius and much of this matter seems to reach the black hole.  Theoretical arguments suggest that the establishment of a powerful wind requires some process to transfer energy from the inflowing gas to the outflow, the nature of which is not understood \citep{shakura73,blandford99}.  In the absence of such a mechanism, there are self-consistent solutions in which the strong outflow is replaced by a gentle circulation or ``breeze" \citep{begelman12a}, in which case supplied gas could accumulate and hyperaccretion would be possible under a wider range of conditions.  Thus, our proposed criterion for hyperaccretion represents a conservative view of the process.

According to our adopted view, hyperaccretion commences only if matter crossing into the black hole sphere of influence has a small enough specific angular momentum compared to the Keplerian value, typically a few percent or less.  While this can be a stringent constraint, it is relaxed considerably for relatively small black holes in protogalactic halos with relatively large velocity dispersions (typically, more massive halos).  This condition is most readily met when the growth of the black hole has lagged behind the $M-\sigma$ relation for its host bulge \citep{ferrarese00,gebhardt00,tremaine02}.  Nevertheless, the likelihood of meeting the angular momentum condition depends on the outer boundary conditions for the mass supply, which we leave for later investigations.  

Once this initial angular momentum condition is met, matter begins to accumulate in an envelope at a rate slightly lower than the growth rate of the black hole.  As the envelope grows --- in radius as well as mass --- the specific angular momentum in its outer regions increases, which means that the angular momentum constraint for maintaining an episode of hyperaccretion actually weakens with time.  The increasing mass of the envelope is needed in order to store the angular momentum left behind when gas is swallowed by the black hole, until some mechanism is able to remove it.  Barring such a mechanism, we show that the density distribution in the envelope approaches the slope $\sim - 3/2$ characteristic of free-fall and Bondi accretion.  

This enables us to estimate the thermal and radiative properties of the envelope, which must radiate at $\sim L_\edd$.  While the effective temperature of the outer envelope is typically a few thousand K, the radiation is thermalized only for $\dot M \gta 10^3 \dot M_\edd$; hyperaccreting black holes in this regime would resemble red giants.  For lower values of $\dot M / \dot M_\edd$ the the color temperature rapidly increases, until the envelope becomes a hard X-ray source for $\dot M \lta$ a few hundred $\dot M_\edd$.  

In addition to the isotropic emission from the envelope, we expect hyperaccreting MBHs to produce jets that carry most of the accretion luminosity, which could be orders of magnitude larger than the envelope emission.  While the nature of the jet production mechanism is unclear, and may be different in different situations (e.g., magnetic propulsion vs.~driving by radiation pressure), analogy with hyperaccreting TDEs \citep{kara16} suggests that a subset of hyperaccreting MBHs might be most readily detectable though intense, geometrically beamed X-ray emission.  Such sources would be rare, however, not only because of beaming but also because we would expect most hyperaccreting MBHs to have relatively small masses, which implies that their lifetimes are short and their numbers relatively small.

Our investigation suggests an explanation for why very few if any hyperaccreting MBHs have been identified: truly hyperaccreting sources would not resemble AGN.  Either they would be intrinsically X-ray weak because the temperature of the envelope (not a standard accretion disk) is relatively low ($\lta 10^4$ K), or they would be heavily obscured, extremely hard X-ray sources.  Given their low masses, envelope emission at $L_\edd$ would be hard to pick out at high redshifts.  It is therefore not surprising that standard observational strategies have not detected such sources.  Probably, the best hope would be to detect intense X-ray beams from rare sources pointing at us, which could have quasar-like fluxes; this could provide an exciting window into the early growth of supermassive black holes.

\section*{Acknowledgements}
MCB acknowledges support from NASA Astrophysics Theory Program grants NNX14AB37G and NNX14AB42G
and NSF grant AST-1411879, and thanks the Institut d'Astrophysique de Paris and the Institut Lagrange de Paris for their hospitality and support.  MV acknowledges funding from the European Research Council under the European Community's Seventh Framework Programme (FP7/2007-2013 Grant Agreement no.\ 614199, project ``BLACK'').  

\bibliographystyle{mn2e}
\bibliography{../../biblio}

\end{document}